\pdfoutput=1
\documentclass[aps,prb,reprint,amsmath,amssymb,superscriptaddress]{revtex4-1}
\usepackage{bm}
\usepackage{graphicx}
\usepackage[colorlinks]{hyperref}
\usepackage{mathrsfs}
\usepackage{times}
\usepackage[dvipsnames]{xcolor}
\usepackage{float}
\usepackage{physics}

\newcommand{\ocite}[1]{[\onlinecite{#1}]}

\begin{document}

\newcommand{\titleinfo}{Dynamical hadron formation in long-range interacting quantum spin chains}

\title{\titleinfo}
\author{Joseph Vovrosh}
\affiliation{Blackett Laboratory, Imperial College London, London SW7 2AZ, United Kingdom}

\author{Rick Mukherjee}
\affiliation{Blackett Laboratory, Imperial College London, London SW7 2AZ, United Kingdom}

\author{Alvise Bastianello}
\affiliation{Department of Physics and Institute for Advanced Study, Technical University of Munich, 85748 Garching, Germany}
\affiliation{Munich Center for Quantum Science and Technology (MCQST), Schellingstr. 4, D-80799 M{\"u}nchen, Germany}

\author{Johannes Knolle}
\affiliation{Department of Physics TQM, Technical University of Munich, 85748 Garching, Germany}
\affiliation{Munich Center for Quantum Science and Technology (MCQST), Schellingstr. 4, D-80799 M{\"u}nchen, Germany}
\affiliation{Blackett Laboratory, Imperial College London, London SW7 2AZ, United Kingdom}

\begin{abstract}
The study of confinement in quantum spin chains has seen a large surge of interest in recent years. It is not only important for understanding a range of effective one-dimensional condensed matter realizations, but also shares some of the non-perturbative physics with quantum chromodynamics (QCD) which makes it a prime target for current quantum simulation efforts. In analogy with QCD, the confinement-induced two-particle boundstates that appear in these models are dubbed mesons. Here, we study scattering events due to meson collisions in a quantum spin chain with long-range interactions such that two mesons have an extended interaction. We show how novel hadronic boundstates, e.g. with four constituent particles akin to \textit{tetraquarks}, may form dynamically in fusion events. In a natural collision their signal is weak as elastic meson scattering dominates. However, we propose two controllable protocols which allow for a clear observation of dynamical hadron formation. We discuss how this physics can be simulated in trapped ion or Rydberg atom set-ups.

\end{abstract}

\maketitle


\section{Introduction}
The formation of hadrons in nature has its origin in the peculiar potential energy of the constituent elementary particles, i.e. the pairwise interaction energy of quarks grows indefinitely for increasing separation. This phenomenon, called confinement, is the key feature of quantum chromodynamics (QCD) \ocite{barger2018collider,greensite2011introduction}; and because of its non-perturbative nature, the QCD quantum many-body problem is one of the prime targets of recent quantum simulation efforts \ocite{banuls2020review}. A time-honoured way of studying interacting quantum particles are controlled scattering experiments between different types of constituents of matter, most famously at the large hadron colliders \ocite{doi:10.1146/annurev-nucl-101917-020852, doi:10.1146/annurev-nucl-102115-044812}.

Confinement is also important in condensed matter physics \ocite{PhysRevD.18.1259,Delfino_1996,Delfino_2006} and can be observed, for example, in quantum spin chains \ocite{lake2010confinement,coldea2010quantum,kormos2017real,liu2019confined,rutkevich2010weak,kjall2011bound}, whose elementary excitations are meson-like boundstates of domain-walls. Of course, these one-dimensional (1D) models are much simpler than the full SU$(3)$ gauge theory of QCD, but they do share some of the underlying physics. Thus, basic 1D quantum spin chains may be a first step in simulating a full treatment of the strong force. Furthermore, the simulation of quantum spin chain models is within the capabilities of current quantum simulators, with promising first results \ocite{tan2021domain,vovrosh2021confinement, PhysRevE.104.035309, vidal2004efficient,banuls2020simulating,martinez2016real,Bernien2017,Surace2020}.

Confinement has been shown to lead to exotic non-equilibrium dynamics manifest in a suppression of transport \ocite{kormos2017real,mazza2019suppression,PhysRevB.102.041118} and entanglement spreading \ocite{Scopa2022}, a long lifetime of the metastable false vacuum \ocite{verdel2020real,lagnese2021false,Collura2021, Pomponio_2022} and exotic prethermal phases \ocite{Birnkammer2022,PhysRevB.96.134427,PhysRevB.95.024302}.
Recent efforts have been made in understanding scattering events among mesons \ocite{karpov2020spatiotemporal, milsted2020collisions,surace2021scattering} in the short-range Ising model with both transverse and longitudinal fields.
Despite the rich scattering phenomenology, this setup does not host composite excitations beyond the two-quark mesons. Therefore, observing deep inelastic scattering with exotic particle formation needs a richer microscopic dynamics, as it can be realized, for example, by introducing heavy impurities in the short-ranged Ising chain \ocite{vovrosh2021confinement2}.

\begin{figure}
    \centering
    \includegraphics[width=80mm]{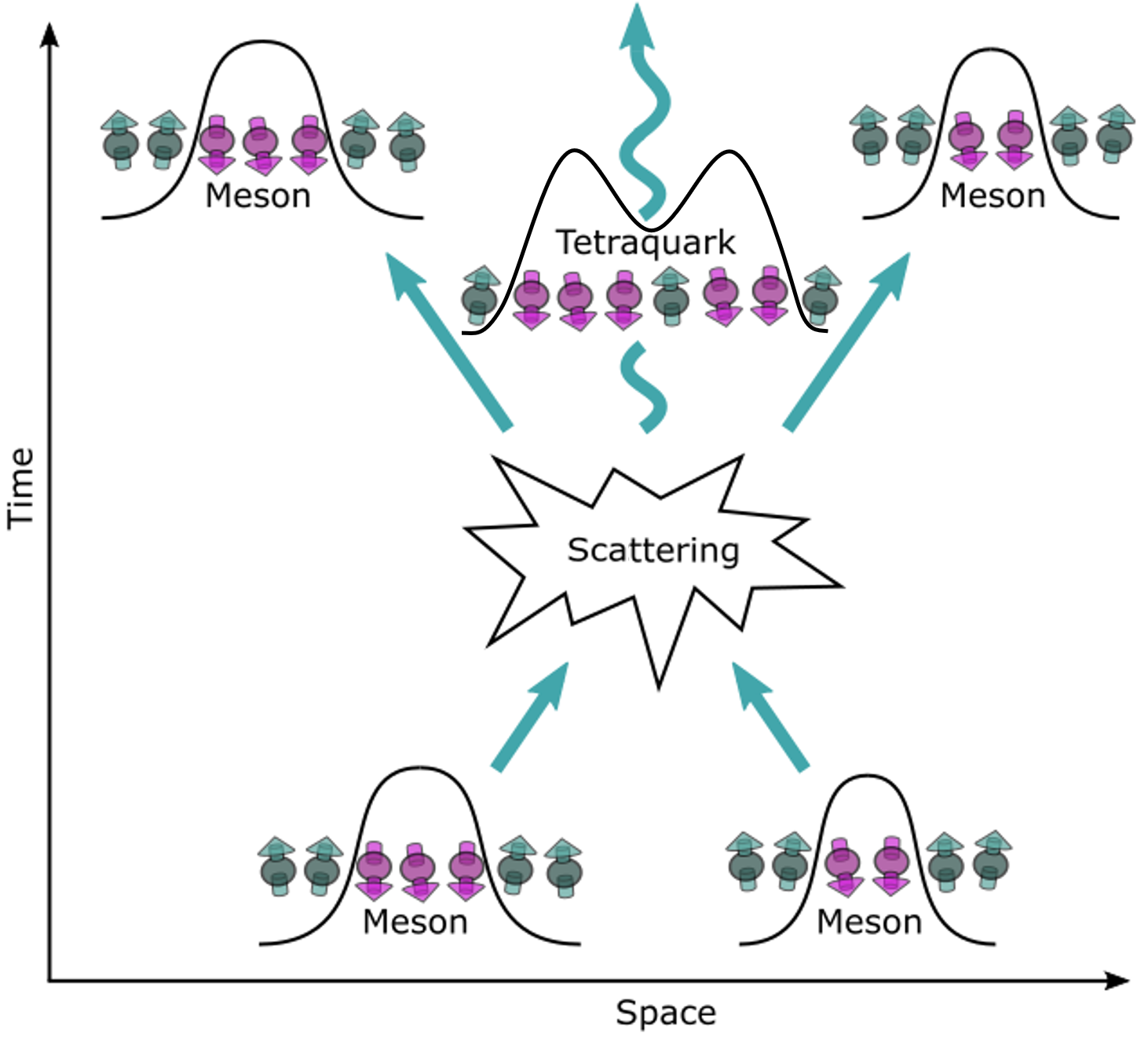}
    \caption{A schematic of a scattering event between two wavepackets of confinement-induced mesonic (two-domain-wall) states. Apart from the elastic meson reflection, the long-range interaction between the composite particles may also lead to the formation of multi-particle boundstates, here in the form of a metastable four-domain-wall state akin to a tetraquark.}
    \label{fig_frontpanel}
\end{figure}

In this work, we investigate another mechanism that can lead to dynamical hadron formation by addressing the transverse Ising chain with long-range interactions which, in addition to confinement \ocite{liu2019confined}, naturally induces interactions among mesons and hosts multi-meson boundstates, akin to hadrons.

We will show that the presence of long-range interactions between mesons can result in the formation of long-lived hadronic metastable boundstates, which can form dynamically in scattering events between fundamental mesons as depicted in Fig. \ref{fig_frontpanel}.

The resulting metastable boundstates are different in nature from infinitely long-lived hadronic boundstates, which cannot be excited in real time scattering processes.
Hence, natural metastable hadron formation is usually weak and the scattering events are dominated by reflection processes.
To enhance the effect, we propose two different modifications of the scattering protocol. Firstly, by an abrupt dynamical change of an external field one can modify the mesons' kinetic energy and induce a non-trivial overlap between the initial asymptotic states and the hadronic bound-states of the post-quench Hamiltonian. As a result, infinitely long-lived boundstates are created and clearly observable.
Secondly, we consider time-independent Hamiltonians with a modified long-range spin interaction, which results in a non-monotonic meson-meson interaction. As a result, scattering mesons can tunnel to a local minimum of the relative interaction and dynamically form a long-lived metastable boundstate, which again becomes clearly observable in the collision.

Our work is structured as follows. 
In section \ref{sec_1}, we consider the low-energy sector of the Hamiltonian by projecting on the four domain-wall subspace. The reliability of such an approximation is related to the extremely long-lifetime of domain-wall excitations, which scales exponentially in the weak transverse field akin to the short range Ising chain \ocite{PhysRevB.102.041118}. In section \ref{sec_2}, we present simulations of the real time dynamics of a collision between two large mesons and find signatures of metastable tetraquark formation. In section \ref{sec_3}, we then show how a simple abrupt change of the kinetic energy can be utilized to allow tetraquarks to form dynamically in a more controllable setting. In section \ref{sec_4}, we further show a second method of inducing fusion by performing a local alteration of the long-range interactions in the spin chain Hamiltonian. We find that tetraquarks with a long lifetime can form dynamically without the need for a time dependent Hamiltonian. In section \ref{sec_5}, we discuss the experimental feasibility of dynamical hadron formation in current quantum simulator platforms. In particular, we discuss initial state preparation as well as how to implement the modified long-range spin interaction within Rydberg atom and trapped ion systems. Finally, we conclude with a summary and discussion.

\section{Hadrons in the long-range Ising model} 
\label{sec_1}
In this work, we consider the one dimensional Ising spin chain with long-range interactions \begin{equation}
    H = -\sum_{i,r}\frac{J}{r^\alpha}\sigma_i^z\sigma_{i+r}^z  -h\sum_{i}\sigma_i^x 
    \label{Hamiltonian}
\end{equation}
where $i$ labels lattice sites, $J$ gives the overall energy scale (we set $J=1$ throughout this work without loss of generality), $h$ is the transverse field strength and $\sigma_i$ are the Pauli matrices. 
The rich dynamics of the long-ranged Ising model has been recently probed in a trapped ion quantum simulator \ocite{tan2021domain}, showing clear signatures of confined excitations.
We are interested in the regime where the transverse field is weak $h\ll J$. The nature of the excitation is best understood by starting with the classical model obtained for $h=0$: here, the natural excitations are domain-walls in the $z-$magnetization and the exponent $\alpha$ crucially determines their interactions. In the limit of large $\alpha$, the short-ranged Ising chain is recovered. Here, domain-walls are non-interacting objects. As $\alpha$ is reduced, longer ranged domain-wall interactions are induced. In particular, two domain-walls $n$ sites apart feel an attractive potential $V(n)$ of the form \ocite{liu2019confined} 
\begin{equation}
    V(n) = 4n\zeta(\alpha) - 4\sum_{1\le l<n}\sum_{1\le r < l}\frac{1}{r^\alpha},
    \label{domainwallpotential}
\end{equation}
and $\zeta(\alpha)$ is the Riemann zeta function.
For arbitrary $\alpha$, the potential $V(n)$ is a monotonically increasing function of $n$ with a finite maximum $\lim_{n\to \infty}V(n)=V_\text{max}<+\infty$ for $\alpha>2$. In contrast, in the regime $1<\alpha<2$, $V(n)$ is unbounded and domain-walls cannot be pulled infinitely far apart without paying infinite energy, thus showing confinement. Finally, for $\alpha<1$ the potential $V(n)$ diverges, signaling a transition to an infinitely-long-ranged model without well-defined local excitations, therefore we will always consider the regime $\alpha>1$.
The activation of a small transverse field $h$ has two effects: \emph{i)} domain-walls become dynamical objects and can hop along the chain, and \emph{ii)} domain-walls in principle are no longer conserved. Nonetheless, resonances between different particle numbers states are greatly suppressed for small transverse field and can be safely neglected \ocite{PhysRevB.102.041118}. Thus, we can project the dynamics in subspaces with conserved number of domain-walls.

\begin{figure}
    \centering
    \includegraphics{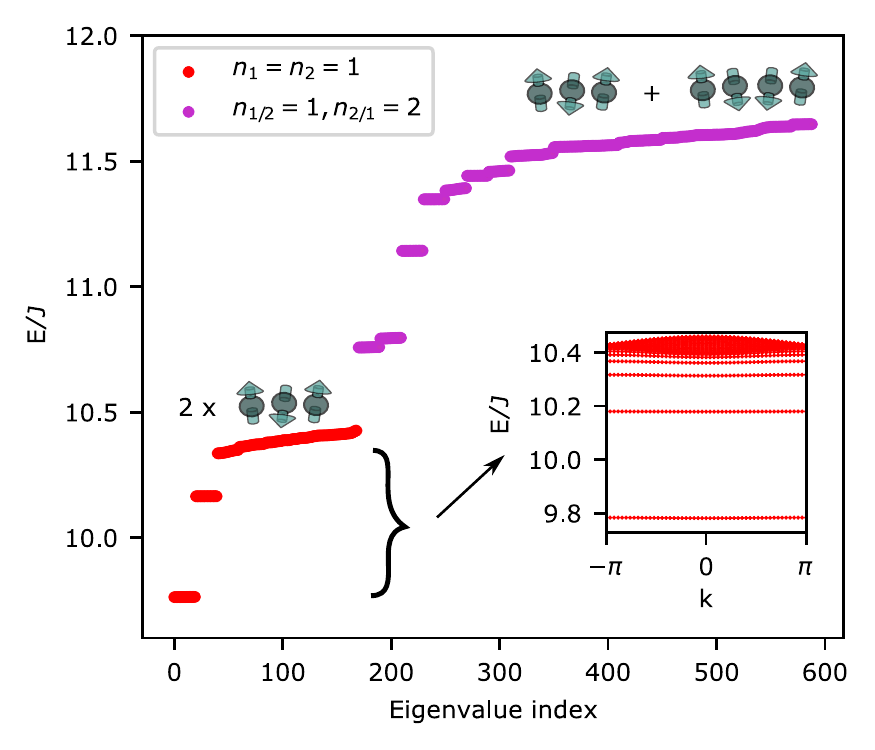}
    \caption{The low energy states of the four domain-wall subspace are shown for $L=20$, $\alpha = 2.6$ and $h=0.1$. The energies strongly depend on the average size of the constituent mesons resulting in large gaps between levels of different meson sizes. A schematic picture of the local spin configurations within each subspace are depicted above the corresponding energy levels. Importantly, due to the long-range interaction the energy within each subspace also depends on the distance between the two mesons, i.e. the closer the mesons are together the lower their energy. The inset shows the energies of the $1$-meson subspace as a function of momentum. The lowest energy levels in this subspace are boundstates of multiple constituent domain-walls.}
    \label{fig_energylevels}
\end{figure}

For example, in the two domain-wall sector one can consider the basis $\ket{j,n} = \ket{\uparrow...\uparrow\downarrow_j...\downarrow\uparrow_{j+n}...\uparrow}$. In this notation, $j$ labels the position of the first domain-wall from the left and $n$ is the relative distance of the second domain-wall with respect to the first. The projected Hamiltonian takes the form \ocite{liu2019confined}
\begin{multline}
    H = \sum_{j,n}V(n)\ket{j,n}\bra{j,n} - h\big[\ket{j+1,n-1}\\+\ket{j-1,n+1}+\ket{j,n+1}+\ket{j,n-1}\big]\bra{j,n} ,
\end{multline}
and can be understood as a kinetic term for each domain-wall with hopping strength $h$ and a potential $V(n)$.
As mentioned above, confinement in a strict sense requires $1<\alpha<2$ such that domain-walls cannot be separated. Nonetheless, also for larger $\alpha$ the two-kink subspace shows the presence of deep boundstates and asymptotic states of freely propagating domain-walls can be very high in energy and extremely difficult to excite. Therefore, as long as asymptotically propagating domain-wall states can be neglected, the dynamics of these deep two-kink boundstates closely resemble those observed in the strictly confined regime, and with a slight abuse of jargon we refer to both as ``mesons". Due to the fact that larger values of $\alpha$ mitigate finite-size corrections caused by the long-range potential, we will mainly focus on the regime $\alpha \sim 2.5$ but stress that similar physics appears for $1<\alpha<2$. 

\begin{figure*}
    \centering
    \includegraphics{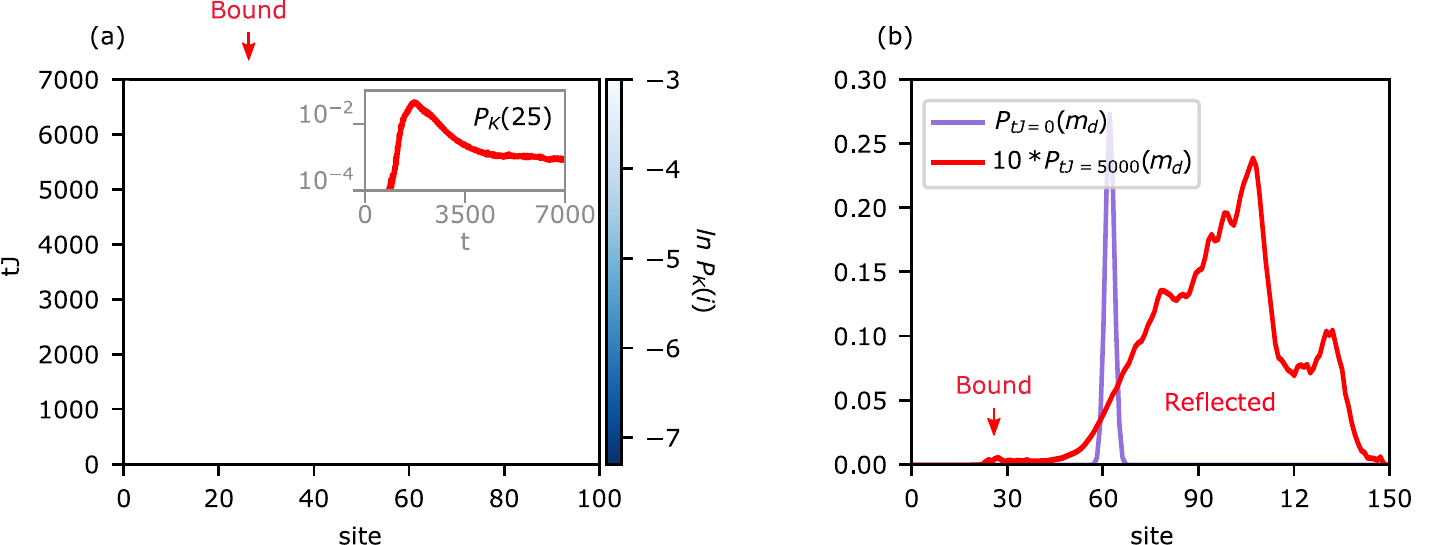}
    \caption{The dynamical formation of a tetraquark through an inelastic collision of large mesons. Here, the Hamiltonian parameters are $L=150$, $\alpha = 2.6$ and $h = 0.1$. The full details of initial wavepacket can be found in the SM \ocite{Note1}. (a) The collision between a mobile meson initialized with a width of 4 sites and a stationary meson with initial width of 10 sites. Although the vast majority of the incoming wavepacket is reflected there is a subtle excitation of a tetraquark that can be seen as the faint vertical lines at sites $\sim 25$. Furthermore, an inset plots the cross section of the time dynamics at site 25 showing the long life time of the tetraquark formed. Note, we choose to plot sites up to 100 as this contains the relevant information for our discussion. (b) The probability distribution of $m_d = j_2+n_2$ at times $tJ=0$, the initial wavepacket, and $tJ=5000$, after the collision event. After the collision, we clearly see that the majority of the wavepacket is reflected, however, there is a subtle peak at site $\sim25$ that corresponds to a bound tetraquark state.}
    \label{fig_fusiondyn}
\end{figure*}

As our goal is the observation hadron-like excitations, we need to consider interactions among mesons. We focus on the dynamics within the four-kinks subspace and use a straightforward extension of the above two domain-wall projection. We consider the basis $\ket{j_1,n_1,j_2,n_2} = \ket{\uparrow...\uparrow\downarrow_{j_1}...\downarrow\uparrow_{j_1+n_1}...\uparrow\downarrow_{j_2}...\downarrow\uparrow_{j_2+n_2}...\uparrow}$, in which the projected Hamiltonian takes the shorthand form
\begin{align}
\begin{split}
    H = & \sum_{j_1,n_1,j_2,n_2}V(n_1)\ket{j_1,n_1,j_2,n_2}\bra{j_1,n_1,j_2,n_2}\\ & + V(n_2)\ket{j_1,n_1,j_2,n_2}\bra{j_1,n_1,j_2,n_2}\\& +
    I(j_1,n_1,j_2,n_2)\ket{j_1,n_1,j_2,n_2}\bra{j_1,n_1,j_2,n_2}\\ & - h \big[\text{hopping terms}\big].
    \label{fourkink}
\end{split}
\end{align}
Here, $I$ can be seen as the meson interaction such that \ocite{Collura2021}
\begin{equation}
    I(j_1,n_1,j_2,n_2) = -4\sum_{j_1< r\le j_1+n_1}\sum_{j_2< s\le j_2+n_2} \frac{1}{(s-r)^\alpha}
\end{equation}
and ``hopping terms" refer to those of the form $\ket{j_1\pm1,n_1\mp,j_2,n_2}\bra{j_1,n_1,j_2,n_2}$ or equivalently for the second meson. Note, the form of the interaction, $I$, depends on the choice of boundary conditions. Here, we consider open boundaries. This effective interaction between mesons falls of as $d^{-\alpha}$ in which $d$ is the meson separation $\sim (j_2 - j_1)$. Thus, similar to domain-walls that are spatially distant from each other, individual mesons which are far apart from each other interact only weakly.

In Fig. \ref{fig_energylevels} we show some of the lower energy levels of the four domain-wall subspace.
We clearly see large energy gaps in correspondence with internal quantum numbers, labeling the energy levels of the two kink-mesons. Additional structure is then provided by boundstates of fundamental mesons and asymptotic scattering states (see inset).
In the case of deep boundstates where the binding energy is much larger than the transverse field $h$, the two-kink mesonic wavefunction is very peaked on integer values of the relative distance, hence the meson has approximately fixed length which is in one-to-one correspondence with the internal energy levels. In this regime, we can pictorially use the meson's size as a good quantum number. Nonetheless, this correspondence is blurred as boundstates become shallower and domain walls can oscillate with internal dynamics.

Before turning to real-time numerical simulations of scattering events, it is useful to further comment on the meson-meson boundstate structure, i.e. our cartoon picture of hadrons.
As shown in Fig. \ref{fig_energylevels}, energy levels corresponding to hadrons are clearly visible in the spectrum, their number and gap with respect to the asymptotic scattering states depends on the ``size" (i.e. the energy) of the binding mesons.
These boundstates are clearly orthogonal to asymptotic scattering states, hence scattering events cannot couple to them. If we wish to observe dynamical hadron formation, we should aim for metastable states arising from asymptotic scattering states whose wavefunction is large when mesons are close, which means that they are qualitatively close to true boundstates.
Heuristically, these states are most likely to be present where the spectrum shows a smooth transition between boundstates and scattering states, i.e. when the energy gap between the two is small. As it is clear from Fig. \ref{fig_energylevels}, this is the case for large mesons.
This picture is also suggested by semiclassical arguments, since for large mesons the relative position of the kinks can oscillate. Therefore, part of the scattering energy of the two incoming mesons can be converted to internal energy of the mesons upon scattering and a boundstate may form. This is not possible for tightly bound mesons, where the relative position of the kinks cannot be changed.
These consideration motivate the use of large mesons in the following natural scattering protocol.

\section{\textbf{Inelastic Collisions of Large Mesons}}
\label{sec_2}

We now consider the Hamiltonian projected in the four-kink subspace and numerically explore scattering events between mesons.
As initial states, we choose Gaussian meson wavepackets  with a well-defined momentum. Furthermore, we fix the average length of each meson to cover a few lattice sites $\sim 4-10$. Details on the wavepacket wavefunction are relegated to the Supplementary Material (SM) \footnote{Supplementary Material at [url] shows details for the derivation of the few-meson subspaces; details on simulations; and details of the WKB approximation for tetraquark decay.}. Simulations in the four kink subspace are challenging for large system sizes. Therefore, we take advantage of global momentum conservation and focus on the sector with  zero total momentum. Using translational invariance we measure the kink coordinates with respect to the leftmost domain-wall which can reduce the Hilbert space dimension by a factor $L$ allowing us to access much larger systems.

In the zero-momentum sector we can then label the Hilbert space by only three variables $\ket{n_1,j_2,n_2}$  as  the first kink can be pinned to $j_1=0$.
As a consequence of this convention, in our real-time simulation the leftmost meson will appear stationary. 

In Fig. \ref{fig_fusiondyn}(a) we show a collision event between a left meson of initial width $n_1=10$ with a right meson with initial width $n_2=4$. We employ the observable $P_K(i) = |\bra{\psi} (1-\sigma^z_i\sigma^z_{i+1}) \ket{\psi} |^2$ which can be seen as the probability that a domain-wall is located on site $i$. 

In Fig. \ref{fig_fusiondyn} we observe that scattering is dominated by elastic reflection events.
However, by plotting the logarithm of the data we are able to highlight subtle details, e.g. a small portion of the wavepacket, $\sim10^{-3}$ remains close to the stationary meson for long time after the collision which  can be seen as a vertical line of intensity at site $\sim25$ Fig. \ref{fig_fusiondyn}(a). 
Next, we study a basic measure of the `distance' between the two mesons $m_d=j_2+n_2-j_1$ . In Fig. \ref{fig_fusiondyn}(b) we plot the probability distribution of $m_d$ before ($tJ=0$) and after the collision ($tJ=5000$). After the collision the distribution has two components. First, the overwhelming probability accounts for reflection events seen as the large hump for sites $>40$. However, there is a second, small peak around site $\sim 25$ which is a signature of our sought-after  metastable tetraquark state.

For all parameter and initial state choices we have explored the signatures of dynamical tetraquark formation are weak as elastic scattering dominates. Therefore, in the following, we propose two simple extensions of the natural meson scattering protocol which allow for an unambiguous observation of dynamical tetraquark formation.

\section{Abrupt change in the transverse field}
\label{sec_3}

In the previous section we have only observed weak signals of dynamically formed tetraquark state. However, true boundstates are clearly present in the spectrum but, as we have already mentioned, these are orthogonal to the scattering states of our wavepacket. In the following we explore the possibility to artificially induce non-trivial overlaps between these two classes of states by dynamical changes in the Hamiltonian. As the most basic example, we consider abrupt changes in the transverse field $h$ at the time of collision $t^*$. We note that the very nature of the so-obtained hadrons is very different from the previously considered metastable states because now true boundstates are excited with an infinitely-long lifetime. 
Another advantage is that within this scenario, we do not necessarily need to target large mesons, which are challenging for experiments. Therefore, we focus on small mesons whose energy levels are approximately in one-to-one correspondence with their length. Hence, we refer to a meson of length $n$ as a $n-$meson.

In the following we focus on the simplest case of $1$-mesons. However, as one could argue that these are just simple magnons, we confirm similar physics for nontrivial $2$-mesons in the SM \ocite{Note1} 

\begin{figure}
    \centering
    \includegraphics{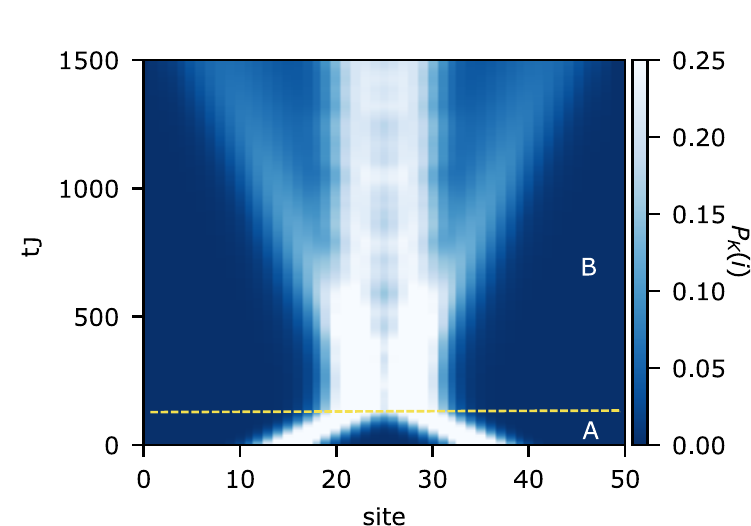}
    \caption{The dynamical formation of a tetraquark with an abrupt change of the transverse field (calculated in the full four domain-wall subspace). Here, the Hamiltonian parameters are $L=50$ and for $\alpha = 2.6$. The full details of the initial wavepacket can be found in the SM \ocite{Note1}. In section A (up to the yellow dashed line) two $1$-meson wavepackets move toward each other with a large kinetic energy (transverse field $h=0.2$). Then at the point of meson collision ($tJ=125$) a abrupt change to a smaller transverse field ($h=0.1$) is performed reducing the kinetic energy in section B. This reduction of kinetic energy can be seen by comparing the fast velocity of the meson wavepackets in section A to the slow velocity in section B. We see that this process induces the formation of a tetraquark with exotic internal dynamics.}
    \label{fig_quenchdyn}
\end{figure}

For a state consisting of two $1-$mesons, a single spin flip naturally leaves the $1-$meson subspace. Thus, in order to get an intuitive feeling about the dynamics within a restricted $1-$meson Hamiltonian we must consider higher order processes. The derivation can be obtained through perturbation theory as detailed in the SM \ocite{Note1}. 
The $1-$meson subspace has the simplified basis $\ket{j_1,j_2} = \ket{\uparrow...\uparrow\downarrow_{j_1}\uparrow...\uparrow\downarrow_{j_2}\uparrow...\uparrow}$ and the Hamiltonian is
\begin{multline}
    H = \sum_{j_1,j_2}h_{j_2-j_1}[\ket{j_1,j_2+1}\bra{j_1,j_2}\\+\ket{j_1-1,j_2}\bra{j_1,j_2}+h.c.]\\ + \big(U_{j_2-j_1}-\frac{4}{n^\alpha}\big)\ket{j_1,j_2}\bra{j_1,j_2} .
    \label{tightmeson}
\end{multline}

We obtain two contributions from second order perturbation theory, an inhomogeneous hopping term, $h_{j_2-j_1}$, and an additional effective interaction, $U_{j_2-j_1}$.
We can take advantage of translational invariance and focus on the sector with global momentum $k$, where we define the relative distance $j=j_2-j_1$ and the momentum-dependent Hamiltonian $H_k$ acting on the states $|k,j\rangle$ is given by
\begin{multline}
    H = \sum_{k,j}2h_{j}\cos{\frac{k}{2}}\big[\ket{k,j+1}\bra{k,j} + h.c\big]\\ +\big(U_j-\frac{4}{j^\alpha}\big)\ket{k,j}\bra{k,j},
    \label{tightmeson2}
\end{multline}
which we can use to calculate the energies as a function of momentum. We note that the energies of the low energy $1-$meson sector
are in very good quantitative agreement with those of the full four domain-wall subspace as well as full exact diagnonalization (ED), see SM \ocite{Note1}.

In the inset of Fig. \ref{fig_energylevels} we show the spectrum of this $1-$meson subspace. One can clearly see that the lowest energy levels are discrete with large gaps between them, which correspond to boundstates of two mesons that reside only a few sites away from each other, i.e. they describe tetraquarks. In addition, there is a continuum of levels at higher energy of far-apart mesons which interact only weakly, i.e they are free to propagate. 

In Fig. \ref{fig_quenchdyn} we implement the proposed protocol. We initialize the $1-$meson Gaussian wavepackets with well-defined momentum and suddenly reduce the transverse field at a time $t^*$ when the scattering takes place (dotted horizontal line in the figure). The main part of the signal remains indeed trapped in a boundstate at the center of the chain, thus realising the desired dynamical hadron-forming event.

\begin{figure}
    \centering
    \includegraphics{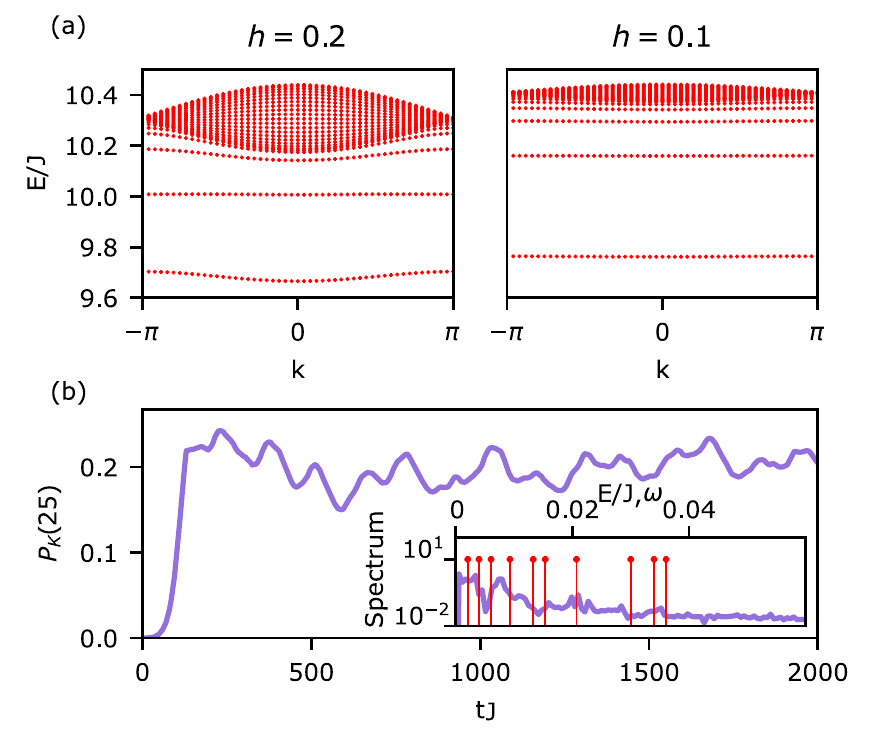}
    \caption{Analysis of oscillations in the fusion event shown in Fig. \ref{fig_quenchdyn}. (a) The energies of the $1-$meson subspace before (left) and after (right) the change in the transverse field at the time of collision. The larger transverse field broadens the continuum of free meson states which corresponds to increasing their kinetic energy. (b) The time dependence of the central site 25 from Fig. \ref{fig_quenchdyn} showing the internal dynamics of the dynamically formed tetraquark. An inset shows the spectrum of the oscillation frequencies observed. Clear agreement is seen between the dominant frequencies observed and differences in the energy levels $E_i$ of the $1-$meson subspace after the change of field to $h=0.1$, these are shown by the vertical red lines.}
    \label{fig_freq}
\end{figure}

In Fig. \ref{fig_freq} we show the dispersion relations of the pre/post-quench Hamiltonians as a function of the total momentum $k$. Boundstates appear as well-separated bands below a high-energy continuum of asymptotic states. Notice that the boundstates for $h=0.1$ have an almost flat dispersion, hence very small velocities, which explains the trapped stationary signal observed in Fig. \ref{fig_quenchdyn}.
Albeit hardly moving, the trapped boundstate is oscillating in time. In Fig. \ref{fig_freq}(b) we show that the oscillating frequencies are indeed compatible with the energy gaps between the boundstate energies of the post-quench Hamiltonian.

\begin{figure*}
    \centering
    \includegraphics{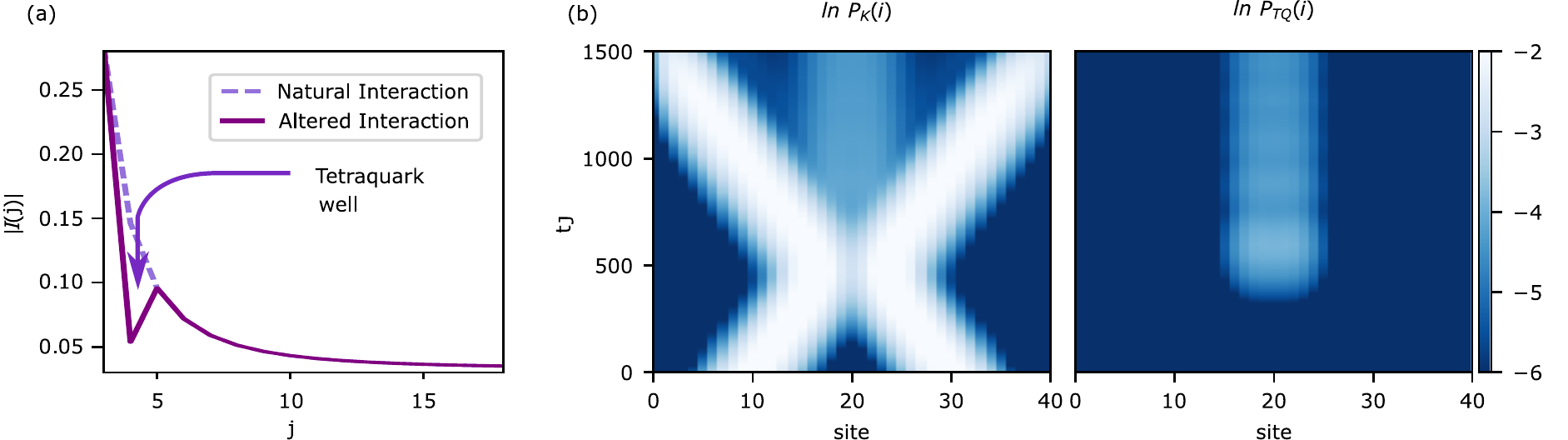}
    \caption{The dynamical formation of a tetraquark from a collision of $1-$mesons using a modified long-range interaction with an additional potential well in the form of Eq. \ref{eq_strongforce} with $d = 4$. Here, the Hamiltonian parameters are $\alpha = 2.6$ and $h = 0.1$. The full details of the initial wavepacket can be found in the SM \ocite{Note1}. Panel (a): The resulting meson-meson interaction is displayed as a function of separation which shows the formation of a tetraquark well. Note, here $I(j) = U_j -\frac{4}{j^\alpha}$. Panel (b): In addition to the main elastic scattering channel, a small portion of the wavepacket quantum tunnels into the tetraquark well forming a long-lived tetraquark state. In the left hand plot we present the real time results for $P_K(i)$ of a meson collision. Here we see that while the majority of the incoming meson wavepackets are reflected, some tunnel into the tetraquark well and persist for long times. In the right hand plot we explicitly show the probability of a tetraquark forming by presenting $P_{TQ}(i)$ for the same collision event.}
    \label{fig_strongforcedyn}
\end{figure*}

\section{Modified long-range interactions}
\label{sec_4}

While the abrupt change of the transverse field protocol  is very efficient in exciting boundstates, it can arguably be regarded as an artificial shortcut to our  goal of observing multi-meson binding in scattering events. Therefore, we now consider possible modifications of the interactions which can enhance the phenomenon without time-dependent changes.

It is again useful to focus on the $1-$meson subspace and consider the relative interaction between two mesons shown in Fig. \ref{fig_strongforcedyn}(a) which is a monotonically decreasing function of the relative distance (dashed line). 
An enhancement of boundstate trapping can then be achieved by modifying the spin interactions to form a potential well.
For example, this can be achieved with the minor modification of the Hamiltonian
\begin{equation}
    H = -\sum_{i,r}\frac{J}{r^\alpha}\sigma_i^z\sigma_{i+r}^z -h\sum_{i}\sigma_j^x +\frac{J}{d^\alpha}\sum_i\sigma_i^z\sigma_{i+d}^z,
    \label{eq_strongforce}
\end{equation}
 inducing a potential well in the relative potential at separation $d$, see solid curve in Fig. \ref{fig_strongforcedyn}(a).
Two colliding mesons will now see a potential barrier of finite width and a quantum tunneling event may occur in the collision.
In the following, we again focus on $1$-mesons for simplicity, but similar boundstate formation can be observed for scattering of larger mesons, see SM \ocite{Note1}.
Physically, this can be understood as increasing the energy of states that have up spins $d$ sites away from each other. In terms of the $1-$meson Hamiltonian this is equivalent to removing the interaction between mesons that are $d$ sites away from each other. 

Fig. \ref{fig_strongforcedyn}(b) shows the effect of the modified interaction for allowing a long-lived tetraquark to form dynamically. Here, we initialized two meson wavepackets with opposite momentum. We observe, as expected, that in the presence of the effective potential well there is a small probability that the mesons tunnels through the potential barrier forming a metastable tetraquark decaying with a lifetime $\tau$. We also show the results of measuring $P_{TQ}(i) = \sum_\mathcal{C}|\psi(j_1,n_1,j_2,n_2)|^2$ where $\mathcal{C}$ is the set of basis states such that there is a domain-wall at the $i^{th}$ site and $j_2-(j_1+n_1) \le d$. This observable can be seen as the probability that a tetraquark is located around site $i$. From this we indeed see that, at the point of collision, a tetraquark forms and has a long lifetime (right panel).

\begin{figure}
    \centering
    \includegraphics{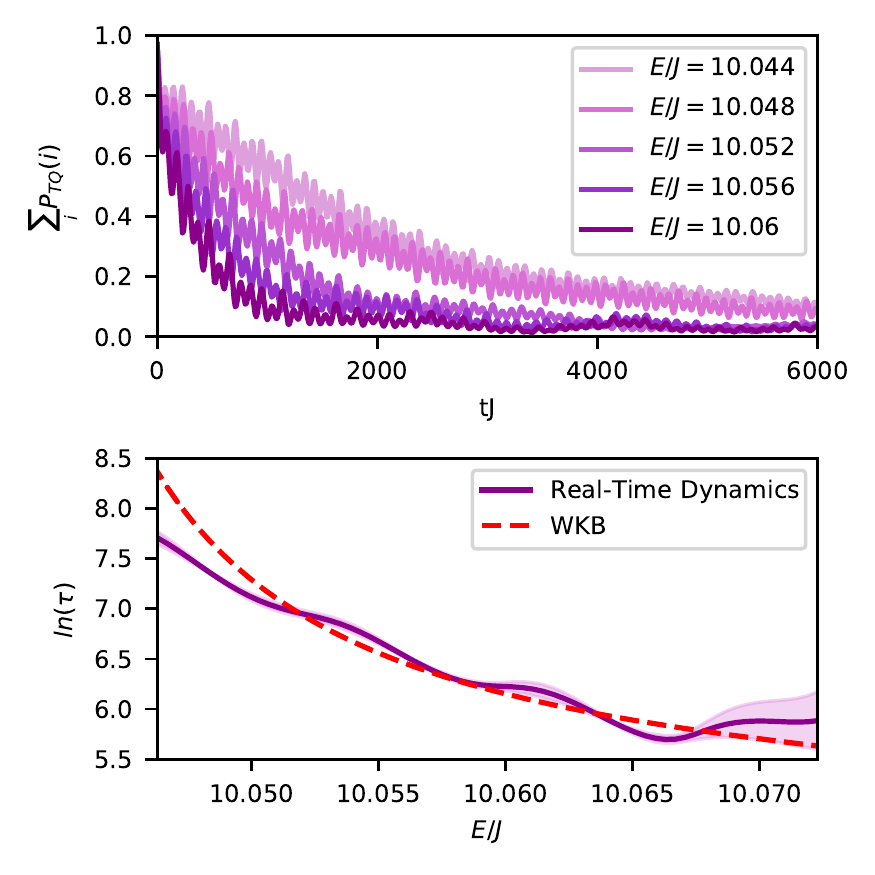}
    \caption{The lifetime of tetraquarks vary as a function of their energy. (a) The decay of tetraquarks for various initial energies. Here, the Hamiltonian parameters are $L=40$, $\alpha=2.6$ and $h = 0.1$. (b) A comparison of the relationship of the the natural logarithm of lifetime, $\ln(\tau)$, of a tetraquark, calculated from the real time simulations as well as the WKB approximation as a function of the initial energy. The WKB approximation agrees well with the lifetimes observed in the real-time dynamics.}
    \label{fig_decay}
\end{figure}

The decay of a metastable boundstate via quantum tunnelling is reminiscent of one of the first paradigms of quantum physics -- $\alpha$-particle decay as observed about a century ago and explained by Gamow's famous semi-classical theory \ocite{griffiths2018introduction}. We can go one step further with our spin chain analogy and compare the decay times of bound tetraquarks for varying energies, which can be achieved through varying the depth of the potential well, $\frac{J}{d^\alpha} \rightarrow \frac{J}{d^\alpha} - \epsilon$. In  Fig. \ref{fig_decay}(a) we plot $\sum_i P_{TQ}(i)$ to observe the decay of states initialized as a tetraquark wavepacket for varying potential well depths. We observe that tetraquarks decay exponentially in time. The lifetime is longer for tetraquarks with lower energy which  is consistent with the larger potential barrier between the bound tetraquark and free meson states. 

Similar to Gamow's theory of $\alpha$-particle decay, we can utilize a WKB approximation to calculate the transmission probability of a tetraquark through the potential barrier. We note that our lattice calculation with a rapidly changing potential is beyond the strict range of applicability of any semi-classical approximation. Nonetheless, we will show decent agreement with our numerical calculations which corroborates the use of spin chain simulators of small size for probing confinement physics. 

We only summarize the results of the WKB calculation with  full details given in the SM \ocite{Note1}. The transmission coefficient of a meson wavepacket is 
\begin{equation}
    T(E) \propto e^{2\int_{x_0}^{x_1}q(x)dx},
\end{equation}
where $x_0$ and $x_1$ are the classical turning points of the potential well and the momentum is
\begin{multline}
    q(x) = \ln\Bigg[\frac{-(U_{n}-\frac{4}{x^\alpha}-E)}{2h_{-1}}+\\\frac{\sqrt{(U_{n}-\frac{4}{x^\alpha}-E)^2-4h_{-1}h_{-1}}}{2h_{-1}}\Bigg].
\end{multline}
One can estimate the probability of emission at any given time by $\frac{2x_1}{v}T$ in which $v$ is the average velocity of the boundstate. Due to the exponential nature of $T$ it gives the dominating contribution to the lifetime $\tau$, which can be predicted within the WKB theory as 
\begin{equation}
  \ln(\tau) \sim C + 2\int_{x_0}^{x_1}q(x)dx
\end{equation}
with a constant $C$ \ocite{griffiths2018introduction}. In Fig. \ref{fig_decay}(b) we show that, by fitting the constant $C$, we see a remarkably good agreement between the WKB theory calculations of the lifetime of tetraquarks and those extracted from the numerical simulations up to the limit of a deep tetraquark well, $\frac{E}{J}<10.045$.

\section{Experimental Feasibility}
\label{sec_5}

A physical realization of our meson scattering protocols  requires the implementation of the one dimensional transverse field Ising model with power-law interactions as described by Eq.~\ref{Hamiltonian}. Although there are several platforms available for the experimental realization of long-range Ising models with hundreds of spins, e.g. with polar molecules \ocite{DipolarNi,Yan}, trapped ions \ocite{britton2012engineered} and Rydberg atoms \ocite{schauss2012observation,Labuhn,Ebadi,Brown}, they are often realized in higher dimensional lattices. However for one dimensional systems, the number of spins  available for quantum simulation range from 3-53 for ions trapped in a Paul trap \ocite{Edwards,Kim,Islam,zhang2017observation} or 20-51 in the case of trapped array of Rydberg atoms in optical tweezers \ocite{Omran, Bernien}, thus reaching the necessary system sizes for our scattering protocols.

In these setups, the many-body state with all spins  down can be naturally realised. Then the initial state preparation of a pair of mesons (domains) located far apart from each other can be achieved through local quenches with help of lasers by performing flips on one or more spins at appropriate sites. The lattice spacing and the width of the laser beam determines the size of the domains which has been demonstrated both for trapped ions \ocite{QuasiparticleJurcevic,martinez2016real,Friis} and Rydberg atoms\ocite{Omran,Labuhn2,Bernien}. For initial states to propagate along the spin chain with well defined momentum requires controlled nearest neighbour spin exchange. For trapped ions, spin exchange between sites is engineered using M\o{}lmer-S\o{}renson type protocols \ocite{QuasiparticleJurcevic}. Effective transitions between $\ket{\uparrow}$ and $\ket{\downarrow}$ are driven with the help of bichromatic laser with frequencies $\omega_{\pm} = \omega_0 \pm \Delta$ that shines upon the ion array, where $\Delta$ is the detuning of the laser field from  the transition frequency $\omega_0$ of the atomic transition $\ket{\uparrow} \leftrightarrow \ket{\downarrow}$. While for Rydberg systems, the spin exchanges are achieved either by resonant dipole-dipole interaction or by off-diagonal van der Waals flip-flop interaction between Rydberg states \ocite{Schempp}. In some case, spin-exchange can also be induced via the standard van der Waals interaction \ocite{Yang}.

Apart from the natural scattering dynamics of the mesons,  we have proposed two other effective routes for the dynamical formation of tetraquarks: (i) by abruptly changing the kinetic energy as depicted in Fig. \ref{fig_quenchdyn}, and by (ii) fusion of mesons induced via a potential well as shown in Fig. \ref{fig_strongforcedyn}(a). Method (i)  can be implemented by abruptly changing the time dependent transverse field. Such type of quench dynamics are commonly executed in both systems, trapped ions \ocite{NonRicherme, tan2021domain}  as well as Rydberg atoms \ocite{schauss2,Keesling}. An advantage of trapped ions is that one can easily tune the strength of the long-range interactions ($1<\alpha<3$) in the Ising model with the help of Raman lasers as recently demonstrated in the  investigation of domain-wall dynamics \ocite{tan2021domain}.

However, the approach (ii) requires a non-trivial modification of the spin-spin interactions to a non-monotonic long-range form. Such exotic potentials may not be obvious for trapped ions, but in the following we sketch how they can be implemented using the highly tunable effective interactions of laser-dressed Rydberg atoms \ocite{Glaetzle1, Gil, Bijnen, Glaetzle2}. A dressed Rydberg atom is primarily a ground state atom that is weakly superposed with an excited state corresponding to a large principal quantum number ($n\sim20-100$) \ocite{Johnson}. The amount and type of the Rydberg character in the dressed superposition state is controlled using  dressing lasers that eventually determine the strength and shape of the effective potential \ocite{Balewski_2014, Mukherjee}. Recent experimental validations of Rydberg-dressed interactions include the  measurement of pairwise interaction between two atoms \ocite{Jau}, many-body Ising interactions \ocite{Zeiher, Borish}, as well as distance selective interactions \ocite{hollerith2021realizing}. 

The key idea for our modified long-range interaction is to construct a spin-1/2 state using two-long lived states ($\ket{g_{\pm}}$) of an atom which can either be a pair of hyperfine ground states or a ground state and a meta-stable state found in alkaline-earth atoms \ocite{Mukherjee}. A pair of lasers (left and right circularly polarized with phases set to zero without loss of generality) drive the atoms from $\ket{g_{\pm}}$ states to Rydberg states $\ket{e_{\pm}}$ off-resonantly with detunings $\Delta_{\pm}$ and Rabi frequencies $\Omega_{\pm}$. The full Hamiltonian in the atomic basis is $H = \sum_i (H^{\text{A}}_i + H^{\text{L}}_i) + \sum_{i<j} H^{\text{vdW}}_{ij}$, where the individual terms are the following
\begin{eqnarray}
H^{\text{A}}_i &=& -\Delta_{+} \ket{e_+}_i\bra{e_+}	-\Delta_{-} \ket{e_-}_i\bra{e_-}, \\	
H^{\text{L}}_i &=& \frac{\Omega_{+}}{2}\ket{g_-}_i\bra{e_+} + \frac{\Omega_{-}}{2}\ket{g_+}_i\bra{e_-}, \\
H^{\text{vdW}}_{ij} &=& V(\mathbf{r}_{ij}) \ket{e_{\alpha}}_i\bra{e_{\alpha}}\otimes\ket{e_{\alpha'}}_j\bra{e_{\alpha'}},
\end{eqnarray}	
where  $V(\mathbf{r}_{ij}) = C_6(\theta,\phi)/r^6_{ij}$ is the van der Waals interaction between Rydberg atoms located at sites $i$ and $j$ and $\alpha, \alpha' \in \{\pm\}$.The effective spin-spin interactions between two atoms can then be obtained with respect to their ground states in the weak coupling limit $\Omega_{\pm} \ll E_{\alpha,\alpha'}$, where $E_{\alpha,\alpha'}$ are the eigenenergies of the relevant atomic systems, $H^{\text{A}}_i + H^{\text{A}}_j + H^{\text{vdW}}_{ij}$. Using perturbation theory, one  derives the effective interactions between the ground states which are expressed in the two-atom basis as follows,
\begin{equation}
H_{\text{eff}} = \sum_{\substack{\alpha,\beta \\ \alpha',\beta'}} 	\tilde{V}^{\alpha,\beta}_{\alpha',\beta'}(\Omega_{\pm},\Delta_{\pm},V(\mathbf{r}_{ij}) )\ket{g_{\alpha}g_{\beta}}_{ij}\bra{g_{\alpha'}g_{\beta'}} .
\end{equation}	
The second order terms in the perturbation theory correspond to light shifts which serve as longitudinal fields for the spin Hamiltonian, while the fourth order terms provide the Ising interaction and transverse field term. The anisotropy in the van der Waals interaction depends on the magnetic quantum numbers and the relative angles $(\theta, \phi)$ between the atoms with respect to the laser.

Overall, our main point is that there is sufficient tunability of the spatial interaction profile in order to engineer a local minimum of the effective interaction similar to the one of Fig. \ref{fig_strongforcedyn}. However, while this effective modified potential between mesons can be achieved through the van der Waals interactions, its asymptotic decay $1/r_{ij}^6$ is not enough to ensure the stability of individual meson states. In order to stabilise them one can exploit the long-range dipole-dipole interactions which is also present. For this purpose, one then needs to consider both levels of the two-level system as Rydberg states for which bound states have indeed been shown to exist \ocite{Letscher}. Alternatively, switching on a weak longitudinal field would also ensure the formation of individual meson states while there mutual interaction would be governed by the longer range potential discussed above.

A detailed quantitative modeling of experimental protocols is beyond the scope of this work, however we provide here some estimate for coherence times. In case of Rydberg-dressed setups, the typical values of Rabi frequencies are in the range of tens of kHz to few MHz while the detunings are an order of magnitude larger. Rydberg states ranging from $n=50-70$ for Rb atoms will have bare lifetimes around $80-130\mu$s \ocite{Saffmanreview}. However as a result of the weak coupling to Rydberg states, the lifetime of the effective two-level system is extended to milliseconds \ocite{hollerith2021realizing}. For such Rydberg states, the interactions are in the order of GHz \ocite{Saffmanreview} and thus for lattice spacings of 0.5-1.5 $\mu$m, one can simulate effective spin-spin interactions which are in the order of few kHz \ocite{Zeiher} to hundred of kHz \ocite{Jau}. Assuming $J=800$ kHz and the time of scattering occurring at $tJ = 100$ (after optimization of the protocol dynamics), coherence times of $0.125$ ms are well within the reach of many-body Rydberg-dressed state lifetimes \ocite{hollerith2021realizing}. Although the dynamical timescales shown in this theoretical work are beyond what has been observed in experiments previously, the use of quantum optimal control theory \ocite{QOC0,QOC1,QOC2} to reduce these timescales is promising especially with the access to a variety of control parameters such as maximising transverse field $h$ while maintaining the domain wall approximation, minimising the initial meson separation, tuning the initial meson size and exponent $\alpha$. Indeed achieving the dynamical timescales and coherence times needed for this work is a challenge for future experiments.

\section{Conclusions}
\label{sec_6}

We have shown that confinement-induced boundstates of many elementary domain-wall excitations exist in the long-range Ising model akin to composite hadronic particles in QCD. We studied the collision of simple two-particle mesonic bound states and found signatures of the dynamical formation of metastable four-domain-wall states in analogy to tetraquarks. However, natural meson collisions are mainly elastic and the signal for natural hadron formation is weak. Therefore, we proposed two alternative protocols to induce dynamical fusion events in the long-range Ising model that are much more controllable and allow for an unambiguous detection of dynamical hadron formation. 

First, we show that an abrupt change in the transverse field at the time of collision results in a strong signal of  tetraquark formation with interesting internal dynamics. Second, we find that a modification of the long-range interaction leads to a tetraquark potential well. Again, we obtain a clear signal of hadron formation and subsequent decay which can be understood via a semi-classical WKB approximation in analogy the to famous example of $\alpha$-particle decay.  

Finally, we argue that all three of these methods, while challenging, are in principle realisable with current quantum simulator set-ups. In particular, we sketch the experimental requirements  for initial state preparation as well as the use of laser-dressed Rydberg atoms for engineering the modified long-range interactions. 

Our work motivates a number of future research questions. For example, we have focused on the limit in which the elementary low energy excitations are well approximated by domain-walls. However, it is well known that for the short-range transverse field Ising part of our Hamiltonion domain-walls continuously evolve into fermionic excitations beyond the small transverse field regime. It would be interesting to gain insight departing from the limit of extremely weak transverse field where, most likely, sharp domain walls will be deformed similarly to what happens in the short-range Ising upon activation of a finite transverse field.

A next step toward the long-time goal of understanding fusion events in full QCD would be the quantum simulation of scattering events with dynamical hadron formation in Hamiltonians that lead to non-mesonic hadrons, such as the $q$-state Potts model with $q>2$ ($q=2$ corresponds to the Ising model) \ocite{Delfino_2008,Lepori_2009,Rutkevich_2015,Liu2020}. Furthermore, one could also consider simplified lattice gauge theories (LGTs). For example, 1D versions of U(1) LGTs have been studied intensively in the past years with many connections to quantum simulation architectures \ocite{banuls2020simulating,martinez2016real}. A prime candidate would be the 1D Schwinger model in which confinement dynamics can be studied efficiently with time-dependent DMRG methods \ocite{magnifico2020real}. In the longer run, similar albeit richer physics is expected in higher dimensional confining LGTs. 

We hope that emulating particle physics scattering experiments  in toy models, whose realisation is feasible for current quantum simulators, can provide a first step towards a better understanding of the fascinating physics of large hadron colliders. As quantum simulators are slowly but surely increasing in size and quality, the simulation of dynamical hadron formation in 1D and beyond would also provide a crucial  benchmark towards achieving a genuine quantum advantage.


\section{Acknowledgements}  
 We are grateful for valuable discussions with Sean Greenaway and Hongzheng Zhao. JV acknowledges the Samsung Advanced Institute of Technology Global Research Partnership and travel support via the Imperial-TUM flagship partnership. The research is part of the Munich Quantum Valley, which is supported by the Bavarian state government with funds from the Hightech Agenda Bayern Plus.
AB acknowledges support from the Deutsche Forschungsgemeinschaft (DFG, German Research Foundation) under Germany's Excellence Strategy -- EXC–2111–390814868.
\newpage


\bibliography{biblio}


\onecolumngrid
\newpage

\setcounter{equation}{0}            
\setcounter{section}{0}    
\setcounter{figure}{0}    
\renewcommand\thesection{\arabic{section}}    
\renewcommand\thesubsection{\arabic{subsection}}    
\renewcommand{\thetable}{S\arabic{table}}
\renewcommand{\theequation}{S\arabic{equation}}
\renewcommand{\thefigure}{S\arabic{figure}}
\setcounter{secnumdepth}{2}
\setcounter{page}{1}

\begin{center}
{\large Supplementary Material \\ 
\titleinfo
}
\end{center}

\section{Derivation of the $1-$meson subspace Hamiltonian}
\begin{figure}[H]
    \centering
    \includegraphics{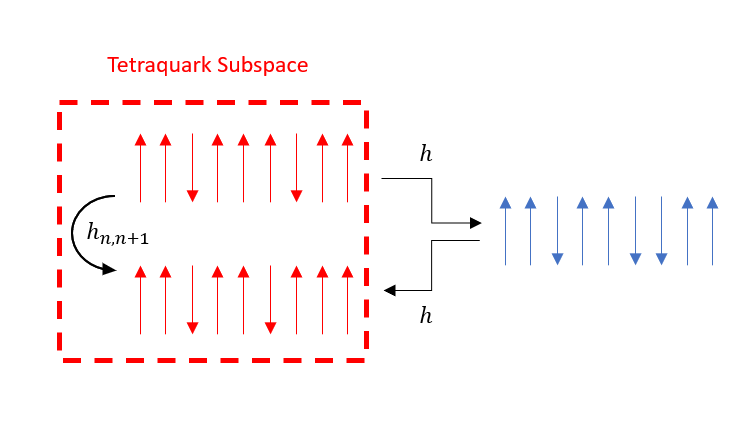}
    \caption{A schematic of the second order hopping process required for a $1-$meson to move.}
    \label{supp_fig1}
\end{figure}

To first order in $h$, projecting the long-range Ising model given in Eq. \ref{fourkink} to the $1-$meson subspace does not include dynamics of the mesons as any spin flip process will leave this subspace. This is depicted in Fig. \ref{supp_fig1}. However, if we consider a second order effect in $h$ we then find terms that allow for these $1-$mesons to hop. The simplest of these are two spin flips which would act similarly to $|j_1, n_1, j_2, n_2\rangle \mapsto |j_1 \pm 1, n_1 \mp 1, j_2, n_2\rangle$ (or the equivalent expression for the other meson). Thus, we must use perturbation theory to accurately describe these $1-$mesons.

\begin{figure}
    \centering
    \includegraphics{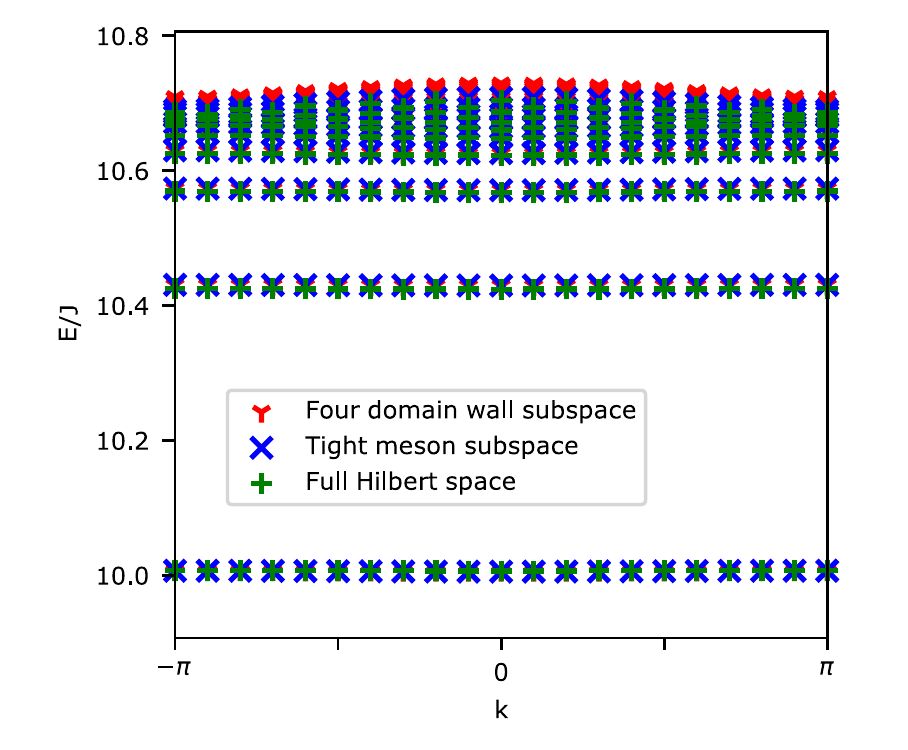}
    \caption{A comparison of $1-$meson energies in the $1-$meson subspace, the four domain-wall subspace as well as the full Hilbert space. Here $L=20$, $h=0.1$ and $\alpha = 2.5$. We see excellent agreement between all three. Note, the energies of the full Hilbert space have been shifted by a constant to account for finite size effects.}
    \label{supp_fig2}
\end{figure}

Our starting point for this projection is the Hamiltonian $H$ given in Eq. \ref{fourkink} which may be factored into a diagonal part $H_0$ and an off-diagonal part $V$ proportional to what we will consider the perturbative parameter $h$, that is $H = H_0 + h V$. Our goal is to find a unitary transformation generated by an operator $S$ such that the expansion $H^{eff} = e^S H e^{-S} \sim H_0 + h^2 M + O(h^3)$ \textit{i.e.} the lowest order hopping events are proportional to $h^2$. $V$ contains the hopping terms of the Hamiltonian; the resulting effective Hamiltonian thus contains only on-site potentials and two-site hopping interactions (plus higher order terms which are absorbed into the $O(h^3)$ factor). This rotation is known as the Schrieffer-Wolff transformation. To proceed, we write down the Baker–Campbell–Hausdorff expansion of $e^S H e^{-S}$ up to second order in $h$:
\begin{equation}
    H^{eff} = e^S H e^{-S} = H_0 + h V + [S, H_0] + h[S, V] + \frac{1}{2}[S, [S, V]] + O(h^3) \ .
\end{equation}
We can eliminate the first order terms by choosing $S$ such that $[S, H_0] = -h V$. The simplest way to do this is element-by-element, enforcing that 
\begin{equation}
    \langle p | S | q \rangle = \frac{ h \langle p| V | q \rangle}{E_p - E_q} \ ,
\end{equation}
where $|p\rangle, |q\rangle$ are eigenvectors of $H_0$ with eigenvalues $E_p, E_q$. With this enforced, the effective Hamiltonian becomes
\begin{equation}
    H^{eff} = H_0 + \frac{h}{2} [S, V] + O(h^3) \ .
\end{equation}
We project this effective Hamiltonian into the $1-$meson subspace and obtain two contributions to the Hamiltonian, a hopping coefficient, $h_{n,n+1}$ in which the $1-$meson moves one site. There is also a second order effect in which a $1-$meson can `hop to itself' through a process in which a neighbouring spin flips and then flips again to leave the initial state unchanged, $h_{n,n}$. In the $1-$meson subspace this can be seen as an effective on site potential. These contributions are given by
\begin{equation}
    h_{j} = \frac{h^2}{2}\bigg[\frac{1}{V(1) - V(2) +\frac{4}{{j+1}^\alpha}} + \frac{1}{V(1) - V(2) +\frac{4}{{j}^\alpha}}\bigg],
\end{equation}
\begin{equation}
    U_{j} = 2h^2\bigg[\frac{1}{V(1) - V(2) +\frac{4}{{j+1}^\alpha}}+\frac{1}{V(1) - V(2) +\frac{4}{{j-1}^\alpha}}\bigg],
\end{equation}
where $j=j_2-j_1$. Given these contributions the $1-$meson subspace can be written as in Eq. \ref{tightmeson}. In Fig. \ref{supp_fig2} we compare the energies of the $1-$meson subspace with the corresponding energies of the four kink subspace as well as the full Hilbert space. Here we see that not only does the $1-$meson subspace have good agreement with the four kink subspace as expected but also with the full Hilbert space.

\section{Meson wavepackets used as initial states in simulations}

The initial state used in the collision of mesons in Fig. \ref{fig_fusiondyn} is given by
\begin{equation}
    \ket{\psi_{t=0}} \propto \sum_{n_1,j_2,n_2} \int dk \: e^{-\frac{(n_1-N_1)^2}{2\sigma_1^2}}e^{ik(j_2+\frac{n_2}{2})} e^{-\frac{(k-q)^2}{2\sigma_k^2}}e^{-\frac{(j_2+n_2/2-c)^2}{2\sigma_c^2}}e^{-\frac{(n_2-N_2)^2}{2\sigma_2^2}}\ket{n_1,j_2,n_2},
\end{equation}
where $q$ is the average momentum of the mobile meson with a standard deviation $\sigma_k$, $c$ is the average centre of mass of the mobile meson with standard deviation $\sigma_c$ and $N_i$ is the average width of the $i^{th}$ meson with standard deviation $\sigma_i$. In particular we use the parameters $q=-1$, $\sigma_k = 0.1$, $\sigma_c = 2$, $N_1 = 10$, $N_2=4$, $\sigma_i = 1$ for $i \in {1,2}$ and $c = 60$.

The initial states used in the collision of mesons in Fig. \ref{fig_quenchdyn} and Fig. \ref{fig_strongforcedyn} are given by
\begin{equation}
    \ket{\psi_{t=0}} \propto \sum_{j_1,j_2} e^{-ikj_1-\frac{(j_1+n-2c_1)^2}{4\sigma^2}} e^{ikj_2-\frac{(j_2+n-2c_2)}{4\sigma^2}}\ket{j_1,n,j_2,n},
    \label{wavepacket}
\end{equation}
where $k$ is the initial momentum of each meson, $c_i$ is the centre of the wavepacket of the $i^{th}$ meson and $\sigma$ is the standard deviation of the wavepacket in real space. In particular the parameters used in Fig. \ref{fig_quenchdyn} are $k = -1.5$, $c_1 = 15$, $c_2 = 35$ and $\sigma = 0.1$. The parameters used in Fig. \ref{fig_strongforcedyn} are $k=-1$, $c_1 = 10$, $c_2 = 30$ and $\sigma = 3$.

\noindent Finally, the tetraquark wavepacket used in the simulations presented in Fig. \ref{fig_decay} is given by 
\begin{equation}
    \ket{\psi_{t=0}} \propto \sum_j e^{-ikj-\frac{(j_1-c)^2}{2\sigma^2}}\ket{j,d}.
    \label{TD_wavepacket}
\end{equation}
where $k$ is the tetraquark momentum (which we set to zero), $c$ is the average position of the first meson and $\sigma$ is the standard deviation of the wavepacket. In particular the parameters used are $d=4$, $k=0$, $c = 18$ and $\sigma = 5$.

\section{Truncation of meson width in simulations}

\begin{figure}[H]
    \centering
    \includegraphics{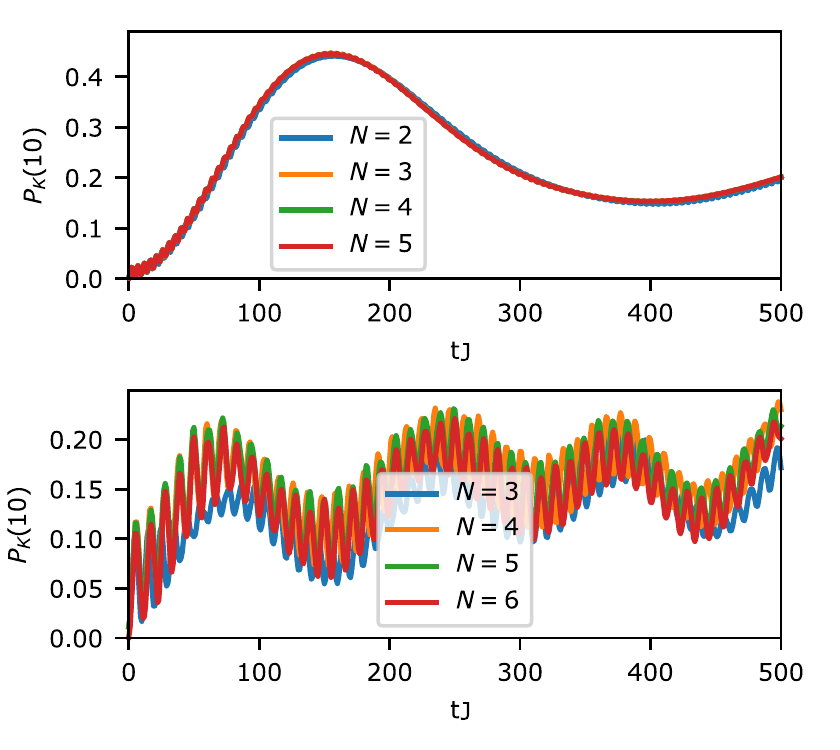}
    \caption{A comparison of meson dynamics for different truncations in meson width, $n_1,n_2 \le N$, for the four domain-wall subspace. Here, $L=20$, $h=0.1$ and $\alpha = 2.5$. The upper panel considers an initial state of two spatially separate $1$-mesons, $\ket{4,1,12,1}$ and the lower panel considers an initial state of two spatially separate $2$-mesons, $\ket{4,2,12,2}$. From this we see that as $N$ increases the dynamics are converge to the what is the dynamics of the full four domain-wall subspace.}
    \label{supp_fig3}
\end{figure}

In order to simulate large system sizes we turned to the four domain-wall subspace described by the Hamiltonian given by Eq. \ref{fourkink}. The size of this subspace grows as $L^4$, here $L$ is system size. This still poses a problem when considering the system sizes required to observe collision events. In order to reduce the computational cost of simulations to quadratic growth, we used a truncation in the meson width, $n_1, n_2 \le N$ for some constant $N$. When considering small mesons, varying the width of a meson vastly changes the energy of the state. Thus, in order to conserve energy, the meson width will only experience small fluctuations. As a result, accurate approximations of the full four domain-wall subspace are achieved by only considering $n_1$ and $n_2$ values below some constant that is at least larger than the meson widths of the initial state. Fig. \ref{supp_fig3} compares time dynamics of mesons for initial states of meson width $1$ and $2$ simulated in the four domain-wall subspace with varying truncations. We clearly see that both $1-$mesons and $2-$mesons are accurately simulated with a truncation of $N \sim 5$. This approximation will naturally become less accurate as the initial state contains larger mesons.

\section{Dynamical hadron formation for larger mesons}

\begin{figure}
    \centering
    \includegraphics{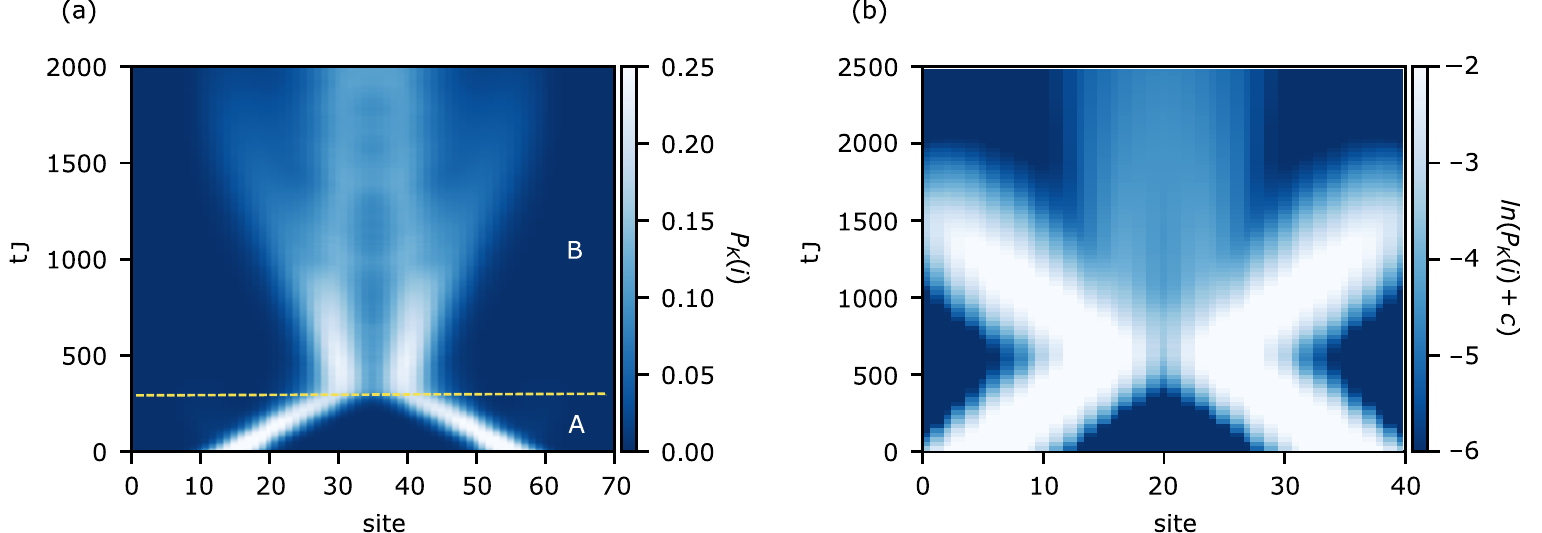}
    \caption{The dynamical formation of a tetraquark from a collision of $2$-mesons. (a) Using an abrupt change of the transverse magnetic field  at $tJ=300$. Here, the Hamiltonian parameters are $\alpha = 2.6$, $h=0.175$ in region A and $h=0.1$ in region B. The wavepacket parameters are The wavepacket parameters are given by $k = -1.5$, $c_1 = 15$, $c_2 = 55$ and $\sigma = 0.1$. Similar results as in Fig. \ref{fig_quenchdyn} are observable except that the oscillations of the tetraquark formed have a larger amplitude as well as slower frequencies, which is expected for a tetraquark formed from mesons with larger masses and non-trivial internal dynamics. (b) Performing a local alteration of the long-range interactions consisting of two terms of the form \ref{eq_strongforce} with $d=4$ and $d=5$. Here, the Hamiltonian parameters are $\alpha = 2.6$, $h=0.1$. The wavepacket parameters are $k=-1$, $c_1 = 10$, $c_2 = 30$ and $\sigma = 3$. We clearly see that in the collision a long lived metastable tetraquark is formed for $2-$mesons.}
    \label{supp_fig4}
\end{figure}

In this section we would like to confirm that our dynamical protocols enabling fusion events in meson collisions is not exclusive for $1-$mesons. One could argue that $1-$mesons are special because they do not have any internal dynamics and are simply equivalent to magnon spin flips. Here, we repeat both the abrupt change in the transverse field as well as the local alteration of the long-range interactions in the Hamiltonian for initial wavepackets composed of $2-$meson. Fig. \ref{supp_fig4} shows the result of the collisions which are very similar as the $1$-meson collision shown in the manuscript. In these we clearly see that the protocols outlined in the this work are not unique to $1-$mesons. The case of the abrupt change in the transverse field also results in continued oscillations localized for a long time around the scattering region for larger mesons. In the case of the induced tetraquark potential well we again see a long lived metastable tetraquark formed during the collision.

\section{WKB Approximation for Tetraquark Transmission Probability}

In this section we derive transmission probability of a meson through a potential barrier of $V(n)$ for a Hamiltonian of the form of Eq. \ref{tightmeson2} with $k=0$ given by
\begin{equation}
H = \sum_n\bigg[J_n\big[\ket{n+1}\bra{n}+\ket{n}\bra{n+1}\big] + V(n)\ket{n}\bra{n}\bigg]
\end{equation}
where, $J_n$ is the site dependent hopping strength. We use the WKB approximation, $\psi(n) = Ne^{\phi(n)}$, in which $\phi$ is a complex function and $N$ is a normalization constant. This leads to the eigenvalue equation
\begin{equation}
Ee^{\phi(n)} = J_ne^{\phi(n+1)}+J_{n-1}e^{\phi(n-1)} + V(n)e^{\phi(n)}
\end{equation}
We make the approximation that $e^{i\phi(n+dx)}\sim e^{\phi(n)+dn\phi'(n)}$ such that $dn =1$ leading to
\begin{equation}
E = J_ne^{\phi'(n)}+J_{n-1}e^{-\phi'(n)} + V(n).
\end{equation}
Given that $\phi(n)$ is smooth, we let $\phi(n) = \int^nq(\epsilon)d\epsilon$. Now
\begin{equation}
(V(n)-E)+J_ne^{q(n)}+J_{n-1}e^{-q(n)} = 0.
\end{equation}
We can solve this quadratic equation in $e^{q(n)}$ to obtain
\begin{equation}
q_{\pm}(n) = \ln\bigg[\frac{-(V(n)-E)\pm\sqrt{(V(n)-E)^2-4J_nJ_{n-1}}}{2J_n}\bigg].
\end{equation}
It turns out that $q_+\sim-q_-$ away from the potential barrier, thus, we can think of $q_+$ as a particle travelling to the right and $q-$ as a particle travelling to the left. Thus, given that $\psi_1$ is the wavefunction in the potential well, $\psi_2$ is wavefunction in the classically forbidden region and $\psi_3$ is the wavefunction after the potential wall we set up the problem as
\begin{equation}
    \begin{aligned}
    \psi_1(n)=e^{\int_{n_0}^nq_+(x)dx}+Re^{\int_{n_0}^nq_-(x)dx}\\
    \psi_2(n)=Ae^{\int_a^nq_+(x)dx}+Be^{\int_a^nq_-(x)dx}\\
    \psi_3(n)=Te^{\int_b^nq_+(x)dx},
    \end{aligned}
\end{equation}
\noindent where $R$ is the reflection coefficient, $T$ is the transmission coefficient and $n=a$ and $n=b$ are the classical turning points.

Physically these are interpreted as $\psi_1$: a particle in the well approaching the potential and a reflection term from a collision with the potential, $\psi_2$: a general solution to the eigenfunction problem above and $\psi_3$: the transmitted particle. We then use the standard procedure that $\psi_1(a)=\psi_2(a)$, $\psi_1'(a)=\psi_2'(a)$, $\psi_2(b)=\psi_3(b)$ and  $\psi_2'(b)=\psi_3'(b)$ to solve for $T$:
\begin{equation}
    T = \frac{e^{\int_{n_0}^a q_-(x)dx}e^{\int_a^b q_+(x)dx} (q^{L}_-(a) - q^{L}_+(a))}{
(q_-(a) - q^{L}_+(a))}
\end{equation}
where the superscript `$L$' indicates if the value of $q_{\pm}$ should be evaluated as the left hand limit of the potential at the discontinuity, $n=a$. Given that $q_-(x)$ is purely imaginary over the interval $n_0<n<a$ and the transmission probability is given by $|T|^2$, we have that the exponential factor of the transmission probability is 
\begin{equation}
    |T|^2 \propto e^{2\int_a^bRe[q_+(x)]dx}.
\end{equation}

\end{document}